\documentclass[fleqn,5p]{elsarticle}

\journal{Applied Energy}

\usepackage{multicol}
\usepackage{amsmath}
\usepackage{amssymb}
\usepackage{mathtools}
\usepackage{caption} 
\usepackage{cuted}
\usepackage{graphicx}
\usepackage{flafter}
\usepackage{placeins}
\usepackage{IEEEtrantools}
\usepackage{textcomp}
\usepackage{hyperref}
\usepackage{color}
\usepackage{float}
\usepackage{booktabs} 
\usepackage{cleveref}
\crefrangelabelformat{equation}{(#3#1#4--#5#2#6)}
\crefname{equation}{}{}
\hypersetup{urlcolor=cyan}

\usepackage{algorithm}
\usepackage{algorithmic}

\DeclareMathOperator*{\argmax}{arg\,max}

\newcommand{\cT}{{\mathcal T}}

\newcommand{\cG}{{\mathcal G}}
\newcommand{\cL}{{\mathcal L}}

\newcommand{\uP}{\underline{P}}
\newcommand{\oP}{\overline{P}}

\newcommand\mydescriptionopt{
	\IEEEsetlabelwidth{$g \in \cG_{\textit{off}}^0$]}
	\IEEEusemathlabelsep
}

\usepackage{flushend} 

\begin{document}

\setlength{\tabcolsep}{6pt}
\setlength{\mathindent}{0pt}

\title{Reinforcement Learning and Mixed-Integer Programming for Power Plant Scheduling in Low Carbon Systems: Comparison and Hybridisation}
\author[imperial]{Cormac O'Malley\corref{cor1}\fnref{fn1}}
\ead{c.omalley19@imperial.ac.uk}
\author[UCL]{Patrick de~Mars\fnref{fn1}}
\ead{patrick.demars@ucl.ac.uk}
\author[UPM]{Luis Badesa}
\ead{luis.badesa@upm.es}
\author[imperial]{Goran Strbac}
\ead{g.strbac@imperial.ac.uk}

\address[imperial]{Imperial College London, South Kensington, London, SW7 2AZ, UK}
\address[UCL]{University College London, Gower St, London, WC1E 6BT, UK}
\address[UPM]{Technical University of Madrid (UPM), Ronda de Valencia 3, 28012 Madrid, Spain}

\fntext[fn1]{Cormac O'Malley and Patrick de~Mars contributed equally to this work.}
\cortext[cor1]{Corresponding author}


\begin{abstract}
Decarbonisation is driving dramatic growth in renewable power generation. This increases uncertainty in the load to be served by power plants and makes their efficient scheduling, known as the unit commitment (UC) problem, more difficult. UC is solved in practice by mixed-integer programming (MIP) methods; however, there is growing interest in emerging data-driven methods including reinforcement learning (RL). In this paper, we extensively test two MIP (deterministic and stochastic) and two RL (model-free and with lookahead) scheduling methods over a large set of test days and problem sizes, for the first time comparing the state-of-the-art of these two approaches on a level playing field. We find that deterministic and stochastic MIP consistently produce lower-cost UC schedules than RL, exhibiting better reliability and scalability with problem size. Average operating costs of RL are more than 2 times larger than stochastic MIP for a 50-generator test case, while the cost is 13 times larger in the worst instance. However, the key strength of RL is the ability to produce solutions practically instantly, irrespective of problem size. We leverage this advantage to produce various initial solutions for warm starting concurrent stochastic MIP solves. By producing several near-optimal solutions simultaneously and then evaluating them using Monte Carlo methods, the differences between the true cost function and the discrete approximation required to formulate the MIP are exploited. The resulting hybrid technique outperforms both the RL and MIP methods individually, reducing total operating costs by 0.3\% on average.

 \end{abstract}

\maketitle

\section*{Nomenclature}

\subsection*{Indices and Sets}
\begin{IEEEdescription}[\mydescriptionopt] 
	\item[$g \in \cG$] Set of thermal generators.
	\item[$g \in \cG_{\textit{on}}^1$] Set of thermal generators which are initially committed (on).
	\item[$g \in \cG_{\textit{off}}^1$] Set of thermal generators which are not initially committed (off).
	\item[$l \in \cL_g$] Piecewise production cost intervals for thermal generator $g$: $1, \ldots, L_g$.
        \item[$n \in N$] Set of quantile-based scenarios for net demand: $1,\ldots,N_n$.
	\item[$r \in \mathcal{R}$] Set of net-demand scenario realisations for schedule testing: $1,\ldots,R_n$.
    \item[$t \in \cT$] Half-hourly time steps: $1, \ldots, T$.
\end{IEEEdescription}

\subsection*{System Parameters}
\begin{IEEEdescription}[\mydescriptionopt]
    \item[$c^{ls}$] Cost of load shed (\$/MWh).
    \item[$c^{wc}$] Cost of wind curtailment (\$/MWh).
    \item[$D_t$] Demand forecast at time $t$ (MW).
    \item[$\eta D_{t,n}$] Demand forecast error at time $t$ for quantile based scenario $n$ (MW).
    \item[$\eta \hat{D}_{t,n}$] Net demand forecast error at time $t$ for quantile based scenario $n$ (MW).
    \item[$q_n$] Quantile of the net demand forecast error for scenario $n$.
    \item[$R_t$]    Spinning reserve requirement at time $t$ (MW).
    \item[$W_t$] Wind forecast at time $t$ (MW).
    \item[$\eta W_{t,n}$] Wind forecast error at time $t$ for quantile based scenario $n$ (MW).
    \item[$\Delta t$] Time-step length (h).
    \item[$\phi_n$] Probability of occurrence for quantile-based scenario $n$.
	
\end{IEEEdescription}

\subsection*{Thermal Generator Parameters}
\begin{IEEEdescription}[\mydescriptionopt]
        \item[$c^{st}_g$] Startup cost of thermal generator $g$ (\$).
	\item[$CP_{g,l}$] Cost of operating at piecewise generation point $l$ for generator $g$ (\$).
    \item[$DT_g$]    Minimum down time for generator $g$ (h).
    \item[$DT^1_g$] Number of time periods the unit has been off prior to the first time period for generator $g$.
    \item[$P_{g,l}$] Power level for piecewise generation point $l$ for generator $g$ (MW); $P^1_g = \uP_g$ and $P^{L_g}_g = \oP_g$.
	\item[$\oP_g$]   Maximum power output for generator $g$ (MW).
	\item[$\uP_g$]   Minimum power output for generator $g$ (MW).
        \item[$U_g^1$]  Initial on/off status for generator $g$, $U_g^1=1$ for $g \in \cG_{\textit{on}}^1$, $U_g^1=0$ for $g \in \cG_{\textit{off}}^1$.
	\item[$UT_g$]    Minimum up time for generator $g$ (h).
	\item[$UT^1_g$] Number of time periods the unit has been on prior to the first time period for generator $g$.
	
\end{IEEEdescription}

\subsection*{Decision Variables (binary)}
\begin{IEEEdescription}[\mydescriptionopt]
	\item[$\boldsymbol{u}_{g,t}$]    Commitment status of thermal generator $g$ at time $t$. 
	\item[$\boldsymbol{v}_{g,t}$]    Startup status of thermal generator $g$ at time $t$. 
	\item[$\boldsymbol{w}_{g,t}$]    Shutdown status of thermal generator $g$ at time $t$.
\end{IEEEdescription}

\subsection*{Decision Variables (continuous)}
\begin{IEEEdescription}[\mydescriptionopt]
	\item[$\boldsymbol{C}_g(\boldsymbol{p}_{g,t,n})$]    Cost of power produced above minimum for thermal generator $g$ at time $t$ at quantile scenario $n$ (\$).
    \item[$\boldsymbol{p}_{g,t,n}$]    Power above minimum for thermal generator $g$ at time $t$ at quantile scenario $n$ (MW).
	\item[$\boldsymbol{p}^{\boldsymbol{ls}}_{t,n}$] Load shed at time $t$ for quantile scenario $n$ (MW).
	\item[$\boldsymbol{p}^{\boldsymbol{wc}}_{t,n}$] Wind curtailment at time $t$ for quantile scenario $n$ (MW).
	\item[$\boldsymbol{r}_{g,t}$]    Spinning reserve provided by thermal generator $g$ at time $t$ (MW).
    \item[$\boldsymbol{\lambda}_{g,t,n,l}$]  Fraction of power from piecewise generation point $l$ for generator $g$ at time $t$ for quantile scenario $n$.
\end{IEEEdescription}

\subsection*{Markov Decision Process}
\begin{IEEEdescription}[\mydescriptionopt]
	\item[$A$] Set of all actions (action space).
	\item[$P(s'|s, a)$] Transition function: probability of transitioning to state $s'$ from state $s$ after taking action $a$.
	\item[$R(s', s, a)$] Reward resulting from transition to state $s'$ from $s$ following action $a$.
        \item[$S$] Set of all states (state space).
	\item[$\pi(a|s)$] Policy: probability of taking action $a$ from state $s$.
	
\end{IEEEdescription}

\section{Introduction}
\label{sec: Introduction}
Unit commitment (UC) is the task of determining the on/off schedules of power plants ahead of time and is one of the fundamental problems in power system optimisation. UC is used for cost minimisation in centrally dispatched systems including North American and Australian markets, and a variety of security assessment and intra-day operational tasks in self-dispatching systems like Great Britain. Decarbonisation is driving increased renewable generation whose power forecasts are intrinsically uncertain. As a result, the efficient scheduling of thermal plants to satisfy demand is becoming increasingly challenging. 

The multi-period UC problem is typically solved using mixed-integer programming (MIP) methods which leverage the powerful algorithms used in off-the-shelf commercial MIP solvers. These achieve near-optimal solutions for most problems but can struggle to solve large problem instances over operational timescales while taking rigorous account of uncertainty \cite{papavasiliou2014applying}. MIP methods are categorised as deterministic \cite{Chen2022} or stochastic \cite{Haberg2019}. Deterministic methods deal with uncertainty by setting an explicit reserve requirement, using past experience and heuristics. Stochastic MIP accounts for uncertainty by minimising expected costs over a finite set of discrete scenarios. Stochastic programs result in lower costs compared to deterministic MIPs \cite{ruiz2009uncertainty,Takriti}, but come at the cost of increased complexity, decreased tractability and the need for explicit knowledge of the forecast uncertainty distributions. As such, system operators exclusively use deterministic formulations in practice. 

A new paradigm of solving UC problems is reinforcement learning (RL), leveraging powerful deep learning techniques to map problem features to solutions. RL agents learn by trial and error through repeated simulation of training cases. Artificial intelligence practitioners advocate for RL as a methodology capable of learning optimal control strategies in a large class of problems, including the operation of electricity networks \cite{rolnick2022tackling, kelly2020reinforcement}. Compared to MIP, RL offers the advantage of off-loading most of the computational cost to a training phase, with the potential for rapid decision making close to dispatch. RL is well-suited to decision making with imperfect information, learning operational strategies to manage uncertainty (such as reserve allocation) without reliance on heuristics. To date, no rigorous comparison of MIP and RL for solving UC problems has been made, partly due to limited crossover and collaboration between RL and power systems research communities.

In theory, RL can accurately approximate any system regardless of its convexity. This is a marked advantage over MIP, which must approximate reality in a convex (often linear) manner. The ability to approximate non-convex systems accurately also gives RL a theoretical advantage over supervised learning techniques for UC. Such methods focus on either speeding up MIP solvers \cite{Xavier2021, Pineda2020, nair2020solving}, or estimating optimal solutions based on a training set of previous MIP solves \cite{Pineda2022}. As such, supervised learning methods require the same convex approximations of non-convex systems as MIP.

Attempts to solve UC problems using RL have mainly focused on model-free methods based on Q-learning \cite{jasmin2009reinforcement, jasmin2016function, navin2019fuzzy, li2019distributed, qin2021solving}, focusing exclusively on small-scale power systems of up to 10 generators. Methods combining RL with tree search have outperformed deterministic MIP approaches on problems of up to 30 generators \cite{DeMars2021, de2022reinforcement}. However, these methods do not reap the benefits of rapid solution generation that is characteristic of conventional model-free RL methods. Furthermore, no existing studies have compared RL with stochastic MIP. 


In this paper, we compare the performance of two RL (model-free and lookahead) and two MIP (deterministic and stochastic) approaches for solving UC problems. The methods are rigorously evaluated on a diverse set of 40 test days for 5 problem sizes (10--50 generators) using an open-source benchmark system formulated in Python. To the best of our knowledge, no previous paper has compared the state-of-the-art of RL and MIP for UC on a level playing field. 

We show that MIP methods significantly outperform RL for all problem sizes in terms of operating costs and exhibit significantly better scaling properties. In addition, the performance of RL is highly unreliable, with large volumes of load shedding for specific problem instances. Whereas the solution times of MIP methods scale super-linearly with problem size, the run time of the RL methods remains roughly constant with number of generators. Rather than discouraging work on RL for UC, the authors hope to focus and encourage future research by presenting a clear-eyed view of its current shortcomings.

One such promising research direction is in the combination of machine learning and MIP techniques \cite{Ruan2021}. An agent learns useful properties of the UC problem during training and, once trained, can generate multiple feasible solutions rapidly. Leveraging this advantage, we propose a novel hybrid methodology that employs the RL agent to generate multiple warm-start solutions, used to initialise concurrent stochastic MIP solves. Using this approach, we outperform conventional stochastic MIP and a baseline warm-starting technique, reducing the cost of uncertainty by 0.3\%, or \$110,000/yr in the 50 generator system. This exemplifies the opportunities for advancement found at the intersection of RL and MIP, with potential for large absolute savings in large systems such as MISO or PJM, each with over a thousand generators.

The contributions of this paper are: 
\begin{enumerate}
    \item To provide the most extensive comparison to date between RL and MIP for solving UC problems.
    \item To improve the best solution of either method through a novel hybrid methodology, using RL to warm-start the stochastic MIP solver. We show that this method outperforms the generic warm starting method of a powerful commercial solver, by leveraging the UC-specific information encoded in the RL agent. 
    \item A variant of the hybrid method is also proposed, taking advantage of parallel computing resources to generate several candidate solutions to the UC. We show that this approach somewhat compensates the unavoidable approximation of stochastic MIP in modelling uncertainty, leading to cost savings when applied to operate a power grid. 
\end{enumerate}



The paper is organised as follows: Section~\ref{sec: methods1} details the UC problem, followed in Section \ref{sec: methods2} by descriptions of the MIP and RL agents used to solve it. Section~\ref{sec: case studies} compares the methods' efficacy via case study results, while Section~\ref{sec: hybridisation} presents the improved results from hybridising the RL and MIP methods. Finally, Section~\ref{sec: Conclusion} gives the conclusions.

\section{The Unit Commitment Problem}
\label{sec: methods1}

To facilitate the fair comparison of RL and MIP for the UC problem, we present a unified problem setup that enables different scheduling agents to be evaluated in a common environment. We begin by describing the method-agnostic problem setup, designed to reflect real-world objectives closely. We then present a practical simulation environment, which we use to evaluate UC solutions on different scenarios, as well as to train the RL agents on a variety of problem instances.

\subsection{Unit Commitment}
An agent must output a plant commitment schedule at 23:30 for every half-hour time period for the following day. To allow for post-solve feasibility checks and dispatch signals to be sent with sufficient notice, the agent's best solution at 10 minutes is used. The agent aims to produce the schedule that minimises the expected cost of meeting demand. Although the plant commitments are fixed, the dispatch is varied in real-time to accommodate net-demand realisations best. Thus an agent's goal is defined as follows:
\begin{multline}
\label{eq: true cost func}
\min_{\textbf{u}}: \mathbb{E} \Bigg[   \sum_{t \in \mathcal{T}} \Bigg( \Delta t ( c^{ls} \cdotp \boldsymbol{p}^{\boldsymbol{ls}}_{t} + c^{wc} \cdotp \boldsymbol{p}^{\boldsymbol{wc}}_{t} ) +   \\ \sum_{g \in \mathcal{G}} \Big( c^{st}_g \cdotp \boldsymbol{v}_{g,t} + CP_g^1 \boldsymbol{u}_{g,t} + \Delta t \cdotp \boldsymbol{C}_g(\boldsymbol{p}_{g,t}) \Big)  \Bigg)  \Bigg]
\end{multline}
Where $\boldsymbol{C}_g(\boldsymbol{p}_{g,t})$ is a generators piece-wise linear production cost \cite{Sridhar2013}. For generator $g$:
\begin{flalign}
\label{eq: piecewise cost curve}
& \boldsymbol{C}_g(\boldsymbol{p}_{g,t}) =  \sum_{l \in \cL_g} (CP_{g,l} - CP_{g,1}) \boldsymbol{\lambda}_{g,t,l} && \forall g \in \cG, \forall t \in \cT
\end{flalign}
\begin{flalign}
\label{eq: power curve}
    & \boldsymbol{p}_{g,t} = \sum_{l \in \cL_g} (P_{g,l} - P_{g,1}) \boldsymbol{\lambda}_{g,t,l} && \forall g \in \cG, \forall t \in \cT
\end{flalign}
The variables $\boldsymbol{\lambda}_{g,t,l}$ are continuous and constrained between 0 and 1. Their sum must equal the commitment status of the generator:
\begin{flalign}
\label{eq: fractions}
 & \boldsymbol{u}_{g,t} = \sum_{l \in \cL_g} \boldsymbol{\lambda}_{g,t,l} && \forall g \in \cG,  \forall t \in \cT 
\end{flalign}

The agents seek the plant commitment schedules that minimise the expected cost of load shed, wind curtailment and generation for the next day. At each timestep generator constraints \cref{eq: piecewise cost curve,eq: power curve,eq: fractions}, \cref{eq:a1,eq:a2,eq:a3,eq:a4,eq:a5,eq:a6}, and the energy balance must be maintained:
\begin{multline}
\label{eq: stoch energy balance}
\boldsymbol{p}^{\boldsymbol{ls}}_{t} - \boldsymbol{p}^{\boldsymbol{wc}}_{t} + \sum_{g \in \cG} \big(P_g^1 \boldsymbol{u}_{g,t} +  \boldsymbol{p}_{g,t} \big) = D_t - W_t + (\eta D_t - \eta W_t) \cr \forall t \in \cT
\end{multline}
The wind forecast error ($\eta W_t$) is defined as an autoregressive moving average process `ARMA($p,q$)' of the form:
\begin{equation}
    \eta W_t = \sum^{p}_{i=1} \alpha_i \eta W_{t-i} + \sum^{q}_{i=1} \beta_i \epsilon_{t-i} + \epsilon_t
\label{eq: arma}
\end{equation}
Where $\epsilon_t$ is a standard normal variable. The demand forecast error ($\eta D_t$) also follows an ARMA($p,q$) process of the same form but with different values for the $\alpha_i,\beta_i$ parameters.

It is impossible to calculate a closed-form solution of the expectation (\ref{eq: true cost func}) over these stochastic parameters. However, it can be closely approximated by finding the uniformly-weighted sum over a large number ($R_n$) of Monte Carlo generated test scenarios. To evaluate a given plant commitment schedule, the wind and demand forecast error realisations for each scenario are found by sampling the ARMA processes. The generator set-points for that scenario are found by solving the economic dispatch problem: this is a convex optimisation problem that can be rapidly solved using the lambda-iteration method \cite{wood2013power}. This is repeated $R_n$ times and averaged assuming uniform likelihood:
\begin{multline}
\label{eq: Schedule test cost}
\min_{\textbf{u}}: \sum_{t \in \mathcal{T}} \Bigg( \sum_{g \in \mathcal{G}} \Big(  c^{st}_g \cdotp \boldsymbol{v}_{g,t} +  CP_g^1 \boldsymbol{u}_{g,t} \Big) + \\ \frac{1}{R_n} \sum_{r \in \mathcal{R}} \bigg(   \Delta t \big( c^{ls} \cdotp \boldsymbol{p}^{\boldsymbol{ls}}_{t,r} + c^{wc} \cdotp \boldsymbol{p}^{\boldsymbol{wc}}_{t,r} \big) +   \sum_{g \in \mathcal{G}} \Delta t \cdotp \boldsymbol{C}(\boldsymbol{p}_{g,t,r}) \bigg)  \Bigg)
\end{multline}

\subsection{Simulation Environment} \label{sec: environment}

The calculation of (\ref{eq: Schedule test cost}) is conducted in a dedicated simulation environment for the UC problem, as described in \cite{DeMars2021}. The environment is openly available as a Python package\footnote{\url{https://github.com/pwdemars/rl4uc}}, and can be used to evaluate the quality of solutions represented as a binary matrix $\textbf{u}$ generated by any optimisation technique. In addition, the simulation environment is used to realise the Markov decision process formulation of the UC problem described in Section \ref{sec: RL agent} and train RL agents. 

The environment uses generator specifications given in \cite{kazarlis1996genetic}, defining quadratic fuel cost curves, minimum up/down times, start-up costs and initial generator commitments for $N = 10$ generators. To create larger problem sizes where $N > 10$, we duplicate the generators. To create individual UC problem instances consisting of demand and wind forecasts, we used publicly-available data from National Grid \cite{NG} and Elexon \cite{BMRS}. 806 complete days of national demand and wind generation data for Whitelee wind farm were retrieved, covering dates from 2016--2019 inclusive. 40 case studies were randomly selected to form evaluation case studies used in Section \ref{sec: case studies}, while the remaining days were used as training data for the RL agents. Wind shedding is penalised at a value of \$40/MWh, while load shedding is penalised at \$10,000/MWh.

\section{Scheduling Agents}
\label{sec: methods2}
Here we define the two MIP (deterministic and stochastic) and the two RL (model-free and lookahead) agents used to solve the UC problem presented in the previous Section. We utilise the generator constraints \cref{eq: piecewise cost curve,eq: power curve,eq: fractions} and \cref{eq:a1,eq:a2,eq:a3,eq:a4,eq:a5,eq:a6} defined in \cite{Morales-Espana2013}, which \cite{Knueven2020} concludes are state-of-the-art. The MIP agents are comprised of a cost function and explicit accompanying constraints that are solved for individual problem instances by an off-the-shelf convex solver. The RL agents follow a decision-making policy that converts the environment state variable into an action (e.g. turn generator on/off). During training, the RL agent implicitly learns the UC and generator constraints via repeated interactions with the UC environment.

\subsection{Mixed-Integer Programming Agents}
\label{sec: MIP agent}
Stochastic optimisation finds the decision variables that minimise the expected generation cost over the day in question. However, for (\ref{eq: Schedule test cost}) to accurately approximate the expected cost, $R_n$ capture the low-probability worst-case scenarios when load shedding may occur at 500 times the cost of generation \cite{Sturt2012,Haberg2019}. This will be accomplished by utilising a large $R_n$. However, this can lead to tractability issues because each scenario optimised over introduces a full additional set of the continuous decision variables and their corresponding constraints.

To solve this issue we use a quantile-based scenario tree \cite{Sturt2012} to accurately approximate the large Monte Carlo generated tree of (\ref{eq: Schedule test cost}) with many fewer scenarios. This section details the scenario generation methodology and then formally defines the deterministic MIP (D-MIP) and stochastic MIP (S-MIP) agents.

\subsubsection{Scenario Tree}
\label{sec: Scenario Tree}

We use a scenario tree to represent the $R_n$ Monte Carlo generated scenarios in (\ref{eq: Schedule test cost}) with many fewer $N_n$ representative scenarios corresponding to a user-defined set of net demand forecast error (NDFE) quantiles. An advantage of this method is that it allows extreme high and low net demand realisations to be explicitly accounted for in relatively few scenarios, which is important due to the high expense of load shedding due to under-commitment (and to a lesser extent, wind shedding due to over-commitment). This method has been shown to give a good balance between tractability and accuracy \cite{Sturt2012}.

Each scenario's cost is weighted by the probability of that quantile's occurrence ($\phi_n$) within the cost function. Unlike in (\ref{eq: Schedule test cost}), scenario probabilities are not homogeneous. Importantly, the binary commitment variables are shared between scenarios. Thus, only 1 commitment schedule is produced per optimisation, regardless of the number of scenarios.

To find one NDFE sample, the wind and demand forecast errors are sampled to give $\eta W_{t,s},\eta D_{t,s} \ \forall t$, where $s$ is the sample number index. These two time series are then added to that day's mean forecasts and clipped to be larger than zero as negative wind or demand has no physical meaning: 
\begin{flalign}
     & D_{t} + \eta D'_{t,s} = \max[0, D_{t} + \eta D_{t,s} ] && \forall t \in \cT
\end{flalign}
\begin{flalign}
     & W_{t} + \eta W'_{t,s} = \max[0, W_{t} + \eta W_{t,s} ] && \forall t \in \cT
\end{flalign}
One net demand realisation is found by subtracting the wind forecast error from demand forecast error:
\begin{flalign}
     & \eta \hat{D}_{t,s} = \eta D'_{t,s} - \eta W'_{t,s} && \forall t \in \cT
\end{flalign}

Within the power balance constraints of each quantile-based scenario, the NDFE values ($\eta \hat{D}_{t,n} \ \forall t$) are equal to the value at which there is a  $q_n \times 10^2 \%$ probability of a lower NDFE realisation occurring. These values are found from $\eta \hat{D}_{t}$'s empirical distribution. This is found by sampling $\eta \hat{D}_{t} \ \forall t$ a large number of times ($\approx100,000$). Collecting these samples of $\eta \hat{D}_{t,s}$ gives the empirical distribution of $\eta \hat{D}_{t}$. 

\begin{figure}[t]
    \centering
    \includegraphics[width=0.95\linewidth]{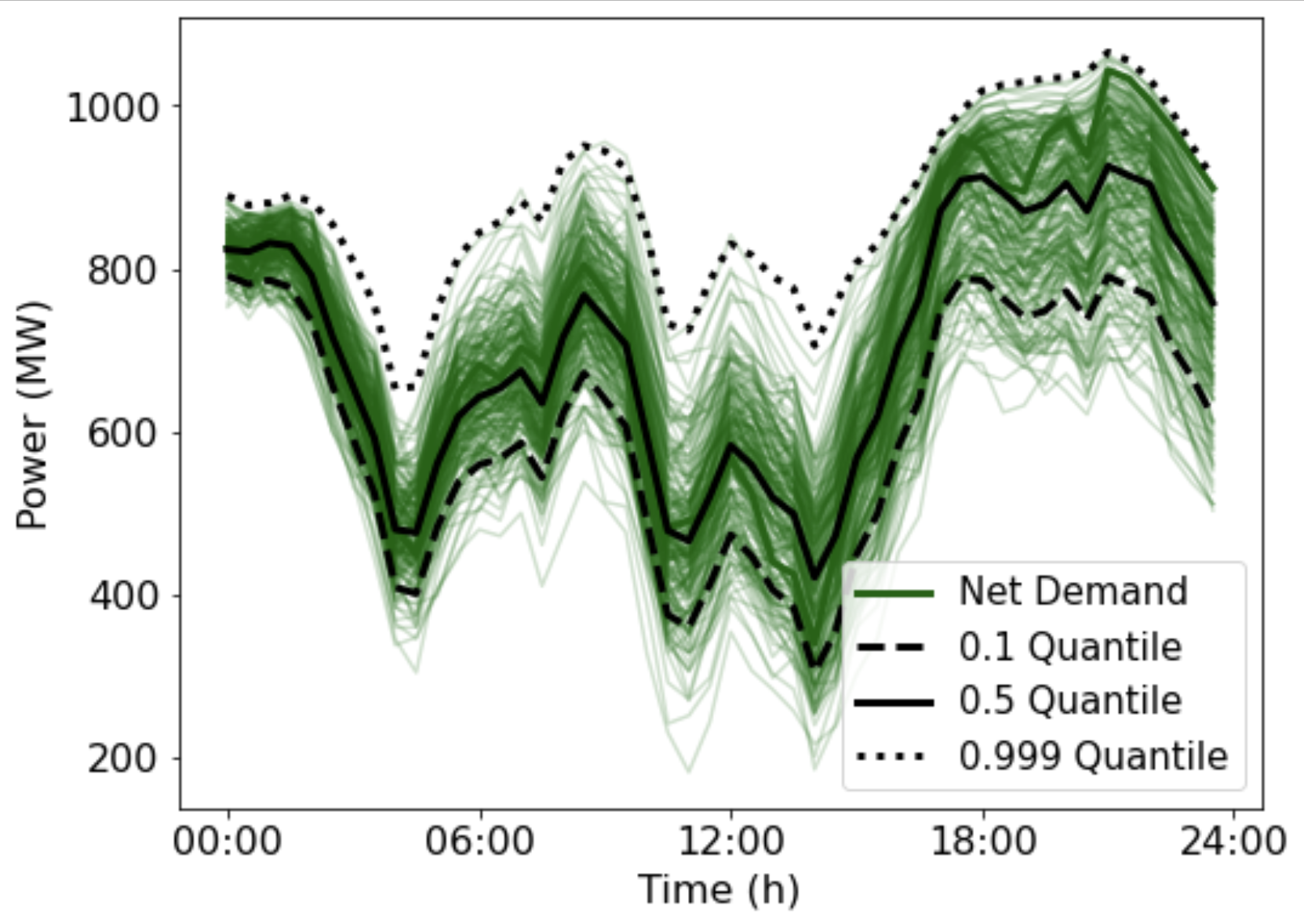}
    \caption{Illustration of how quantile-based scenarios can be used to approximate the empirical net demand forecast distribution to reduce MIP size and, therefore, increase tractability. Plotted are 200 net demand forecast realisations, approximated by 3 scenarios corresponding to the quantiles: ($q_1=0.1,q_2 = 0.5,q_3 = 0.999$).}
    \label{fig: quantile scenarios}
\end{figure}

Fig. \ref{fig: quantile scenarios} plots 200 such scenarios added to that day's net demand forecast. Figure \ref{fig: quantile scenarios} also plots the quantile scenarios: ($q_1=0.1,q_2 = 0.5,q_3 = 0.999$). The $q_1$ corresponds to the bottom scenario, where there is a 10\% chance that the day's net demand will be less than this timeseries. Similarly, for the $q_3$ scenario at the top, only 1 in 1000 net demand realisations will exceed the $q_3$ scenario at each time-step. 

As well as assigning appropriate NDFE values for each quantile-based scenario, the probability  of each scenario's occurrence ($\phi_n$) must also be found. These probabilities are functions of quantile choice and are detailed in \ref{appendix_UC} through eqs.~(\ref{scenario probability}). 

Thus with the probability of each scenario's occurrence known $(\phi_n \ \forall n)$, as well as the half-hourly net demand forecast errors $(\eta \hat{D}_{t,n} \ \forall t , \forall n)$, the updated cost function and demand balance constraints are:

\begin{multline}
\label{eq: scenario tree cost func}
\min_{\textbf{u}}: \sum_{t \in \mathcal{T}} \Bigg( \sum_{g \in \mathcal{G}}  c^{st}_g \cdotp \boldsymbol{v}_{g,t} +  CP_g^1 \boldsymbol{u}_{g,t}  + \\ \sum_{n \in \mathcal{N}} \phi_n \bigg( \Delta t \big( c^{ls} \cdotp \boldsymbol{p}^{\boldsymbol{ls}}_{t,n} + c^{wc} \cdotp \boldsymbol{p}^{\boldsymbol{wc}}_{t,n} \big) +   \sum_{g \in \mathcal{G}}  \Delta t \cdotp \boldsymbol{\hat{C}}_g(\boldsymbol{p}_{g,t,n}) \bigg)  \Bigg)
\end{multline}

\begin{multline}
\label{eq: determ energy balance}
\boldsymbol{p}^{\boldsymbol{ls}}_{t,n} - \boldsymbol{p}^{\boldsymbol{wc}}_{t,n} + \sum_{g \in \cG} \big(P_g^1 \boldsymbol{u}_{g,t} +  \boldsymbol{p}_{g,t,n} \big) = D_t - W_t + \eta \hat{D}_{t,n} \cr \forall t \in \cT, \forall n \in \mathcal{N}
\end{multline}

\subsubsection{Stochastic and Deterministic MIP Agents}

The S-MIP agent is defined as cost function (\ref{eq: scenario tree cost func}) along with constraints \cref{eq:a1,eq:a2,eq:a3,eq:a4,eq:a5,eq:a6}, \cref{eq: piecewise cost curve,eq: power curve,eq: fractions,eq: determ energy balance}. Thirteen net demand forecast scenarios are used, with corresponding quantiles of: [0.002, 0.01, 0.1, 0.3, 0.5, 0.6, 0.75, 0.9, 0.95, 0.99, 0.993, 0.9967, 0.999]. The distribution is discretized more finely over the high quantile range, as these represent higher than expected net demand which can significantly impact the cost function if load-shedding occurs.

By explicitly considering load-shed as a recourse action, and then taking actions to minimise the expected operating costs, the S-MIP implicitly optimises the reserve margin. However, almost all real-life implementations of MIP for UC are deterministic, where the reserve margin is explicitly set.

The formulation can accommodate a deterministic reserve requirement to define the D-MIP agent. This is done by using only one scenario, with zero net demand forecast error ($N_n = 1, \ \phi_1 = 1,\ \eta \hat{D}_{t,1} = 0 \ \forall t \in \cT  $), alongside an explicit reserve requirement:

\begin{flalign}
     \label{eq:explicit reserve requirement}
    & \sum_{g \in \cG} \boldsymbol{r}_{g,t} \geq R_t  && \forall t \in \cT
\end{flalign}

\begin{flalign}
    & p_{g,t} + r_{g,t} \leq (\oP_g - \uP_g) u_{g,t} && \forall t \in \cT
\end{flalign}

In this paper the D-MIP agent sets $R_t$ equal to four times the net-demand forecast error standard deviation for all timesteps $t$.

\subsubsection{MIP Implementation Details}

The MIP is formulated in Python using PYOMO \cite{bynum2021pyomo} and solved by Gurobi~9.5.2 \cite{gurobi}. Gurobi implements a branch-and-bound algorithm for solving MIPs, industry standard for all high-performance commercial solvers. The algorithm is summarised in Algorithm~\ref{alg:B and B1}.

\begin{algorithm}[!t]
  \caption{Branch and Bound Algorithm}
 \label{alg:B and B1}
 
 \begin{algorithmic}[1]
 
 \renewcommand{\algorithmicrequire}{\textbf{Input:}}
 \renewcommand{\algorithmicensure}{\textbf{Output:}}
 
 
  \STATE find an initial feasible solution and set the upper bound (UB) to its objective value
 \STATE relax integrality constraints
 \WHILE{time-limit not exceeded and MIP gap $>$ 0\%}
 \STATE solve resultant continuous linear program(s)
 \IF{first iteration}
 \STATE set lower bound (LB) equal to solution's objective value and go to step 12
 \ENDIF
 \IF{solution is feasible and less than UB}
 \STATE UB equal to solution's objective value
 \ENDIF
 \STATE set LB to lowest leaf node objective value
 \STATE find MIP gap: $\frac{UB-LB}{UB} \cdot 100\%$
 \STATE branch problem into sub-problems (nodes) by constraining some intelligently selected binaries to be 0 or 1. Nodes yet to be branched are referred to as `leaves'

 \ENDWHILE

 \end{algorithmic} 
\end{algorithm}

Step 3 involves solving multiple leaf LPs simultaneously, thus it is well suited to parallelisation across multiple threads.

\subsection{Reinforcement Learning Agents}
\label{sec: RL agent} 

In order to use RL to solve the UC problems, we represent the UC problem as a Markov decision process (MDP), using the formulation described in \cite{DeMars2021}. An MDP is defined as a 4-tuple $(S, A, P, R)$ of a state space, action space, transition function and reward function, respectively. At each timestep $t$, corresponding to decision periods in the UC problem, the agent receives a partial observation of the state $s_t \in S$, and chooses an action $a_t \in A$ sampled from a policy $\pi(a_t|s_t)$. A new state $s_{t+1}$ is returned, sampled from the transition function $P(s_{t+1} | s_t, a_t)$ as well as numeric reward $R(s_{t+1}, s_t, a_t)$. The task of RL is to learn the policy $\pi(a|s)$ that maximises the agent's long-run expected sum of rewards.

The MDP for the UC problem is summarised as follows (for a more detailed description, see \cite{DeMars2021}): 

\begin{itemize}
    \item States $S$: current generator up/down times, demand/ wind forecasts, demand/wind forecast errors, current timestep.
    \item Actions $A$: binary commitment decisions for the following timestep.
    \item Transition function $P(s_{t+1} | s_t, a_t)$: update generator up/down times; sample demand/wind forecast errors from ARMA processes; solve economic dispatch problem determining generator setpoints for demand/wind realisations.
    \item Reward function $R(s_{t+1}, s_t, a_t)$: calculate the operating costs inclusive of fuel, startups, wind shedding and load shedding.
\end{itemize}

The agent observes all elements of the state except the demand and wind forecast errors: this is in order to preserve the day-ahead property of the problem.

The MDP is realised in the simulation environment described in Section \ref{sec: environment}. During training, the RL agent randomly samples a day from the training data set and commits generators through actions $a$, aiming to maximise rewards $r$. We apply a log transformation to the rewards during training, dampening the extreme costs of load shedding which can lead agents to adopt overly conservative policies. 

We use proximal policy optimisation (PPO) \cite{schulman2017proximal} to train the agent, as in previous research \cite{de2022reinforcement, DeMars2021}. Actor and critic neural networks are parameterised separately with deep feed-forward neural networks. Since an action $a$ is a bit-string of length $N$ for a system with $N$ generators, the size of the action space is $|A| = 2^N$. As a result, conventional formulations of the actor as a neural network classifier with $|A|$ output nodes are not tractable except for small problem sizes. We formulate the neural network as a binary classifier which sequentially predicts each action bit $a_i$, corresponding to the commitment decision for generator $i$, conditioned on the commitment decisions for previous generators $a_j$ where $j < i$.

\subsubsection{Model-Free and Lookahead RL Agents}

A trained agent can be used to solve unseen test cases by sampling from the policy $\pi(a|s)$ for new states $s$. To improve reliability during testing, the RL agent chooses actions using `$\argmax_{a} \pi(a|s)$'. We call this agent `model-free' RL (RL-MF) as it does not rely on any lookahead strategy in decision-making.

In addition to the RL-MF agent, we also implemented a 1-step lookahead agent, RL-LA. Using the same policy as RL-MF, the RL-LA agent evaluates a subset of probable actions which meet a threshold $\rho$, subject to $0 < \rho < 1$. That is, all actions where $\pi(a|s) \geq \rho$ are explored. In our experiments, we set $\rho = 0.05$ allowing for a maximum of 20 actions to be explored, in addition to the `do nothing' action which keeps all commitment decisions the same at the following timestep. For each of the actions in the subset, the agent evaluates operating costs over 100 scenarios of demand and wind sampled randomly, and chooses the action giving the lowest operating costs overall. 

Due to the evaluation of multiple actions, the computational expense of evaluating the RL-LA agent is significantly larger than RL-MF, which produces solutions almost instantly. The run times of the two optimisation methods are compared in Section \ref{sec: time comparison}. 

\section{Case Studies and Comparison}
\label{sec: case studies}

In Section \ref{sec: methods1} we described the UC problem setup and simulation environment. In Section \ref{sec: methods2} we described the four optimisation agents: D-MIP, S-MIP, RL-MF and RL-LA. In this section, we present for the first time a detailed comparison of these solution methods, evaluating their performance in terms of operating costs and run time. In addition, we provide a qualitative analysis of their characteristics with case studies. The agents are evaluated on problem sizes of 10--50 generators using the environment described in Section \ref{sec: environment}, with 40 test days each.

Simulations were run on a 24-core (48 thread) 2.85GHz AMD EPYC 7443 processor, with the maximum threads per solve constrained to 8 and a time budget of 10 minutes.

\subsection{Cost Comparison} \label{sec: cost comparison}

Expected operating costs of solutions produced by each of the 4 methods are estimated as in (\ref{eq: Schedule test cost}), using $R_n = 5000$ scenarios, thus providing an accurate estimate of the true expected costs (\ref{eq: true cost func}). In addition to the four optimisation agents, we implemented an agent with perfect foresight. The difference between operating costs produced by the optimisation agents and perfect foresight represents the effective cost of uncertainty.

The mean daily operating costs of the four optimisation agents are shown in Table \ref{tab:mean costs}. Across all problem sizes of 10--50 generators, both MIP agents outperform the RL agents. As expected, S-MIP produces lower operating costs than D-MIP, by between 4.1--4.4\%. Furthermore, RL-LA consistently outperforms RL-MF, benefiting from greater foresight and the ability to consider multiple actions. Overall, the S-MIP agent produces the lowest operating costs on all problem sizes. 

\begin{figure*}[t]
    \centering
    \includegraphics[width=0.97\linewidth]{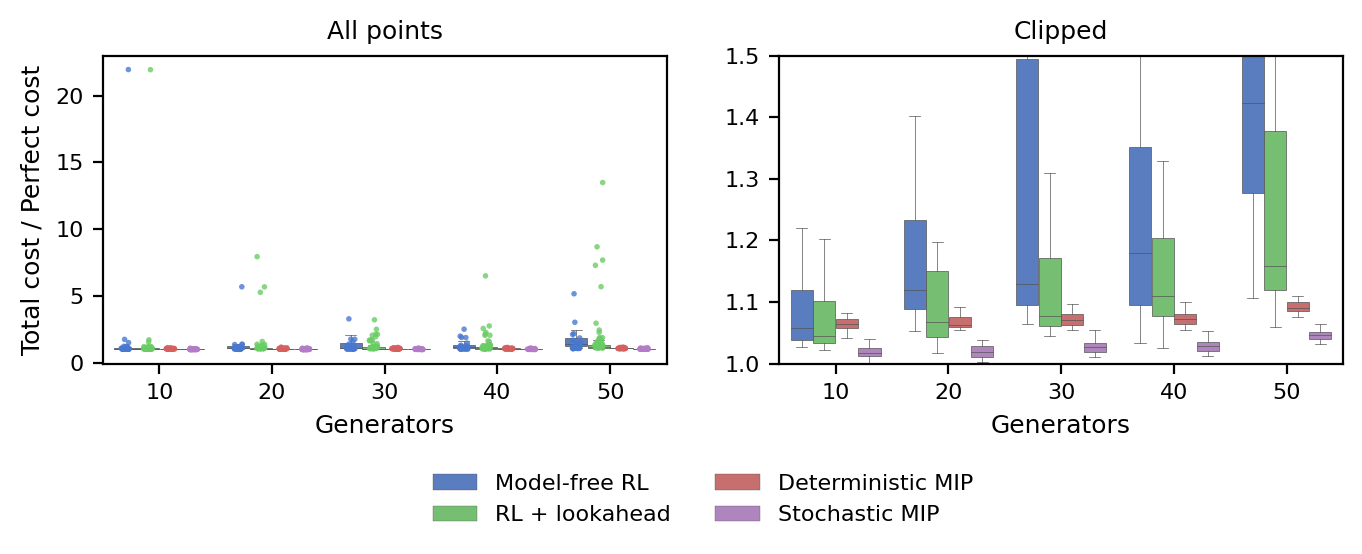}
    \caption{Operating costs relative to perfect foresight costs for the four optimisation agents. The box encompasses the inter-quartile range, with a line at the median. The whiskers extend to the max and min values, but are limited to 1.5x the inter-quartile range. The left-hand plot shows operating costs for all 40 case studies; the right-hand plot is clipped to relative costs $\leq 1.5$. The extreme operating costs of RL-MF and RL-LA in some instances, indicate unreliable performance and difficulties in generalising to unseen problems. However, median operating costs of the RL agents are competitive with D-MIP up to 30 generators.}
    \label{fig:Av Cost}
\end{figure*}

Figure \ref{fig:Av Cost} shows the distribution of mean operating costs over the 40 case studies, normalised against the perfect foresight solutions. The RL-MF and RL-LA results are characterised by greater variability in performance across the different problem instances; for example, RL-MF solutions are up to 21 times more expensive than the perfect foresight costs. This is indicative of the limited ability of the RL agents to generalise effectively to unseen problem instances and the characteristic unreliability of RL as compared with MIP. 

\color{black}

The outliers significantly skew the mean costs for the RL agents; thus, it is also valuable to analyse median operating costs over the 40 problem instances, more clearly visualised in the right-hand plot of Figure \ref{fig:Av Cost}. RL-MF and RL-LA both outperform the D-MIP in the 10 generator system, and RL-LA performs similarly to D-MIP for 20 and 30 generator systems. While the RL agents outperform the D-MIP in many cases up to 30 generators, the solution quality is highly unreliable.

The RL agents show limited ability to scale to larger problem sizes, as shown by significant worsening performance in Figure \ref{fig:Av Cost}. This trend exists to a much lesser extent for both MIP agents due to higher optimality gaps achieved within the 10-minute time budget. The limited ability of the RL agents to scale to large problems is driven by the exponential growth in the size of the action space, which doubles with each additional generator. Intelligently exploring the action space with RL becomes increasingly challenging as the problem size grows, and is not effectively handled by current methods \cite{dulac2015deep}.



\begin{table}[t]
    \centering
    \caption{Sum of mean operating costs for 40 case studies (\$m)}
    \begin{tabular}{lrrrr}
\toprule
Generators &      RL-MF  & RL-LA  & D-MIP &  S-MIP  \\
\midrule
10      &   25.07 &       24.94 &  16.58 &  15.86 \\
20      &   47.58 &       37.32 &  32.97 &  31.52 \\
30      &   61.26 &       54.95 &  49.28 &  47.22 \\
40      &   90.19 &       73.41 &  65.63 &  62.93 \\
50      &  162.26 &      100.66 &  81.94 &  78.62 \\
\bottomrule
\end{tabular}

    \label{tab:mean costs}
\end{table}

\subsection{Solve Time Comparison} \label{sec: time comparison}

\begin{table}[t]
    \centering
    \small
    \caption{Number of Unfinished Solves}
    \begin{tabular}{lrrrrrr}
        & \multicolumn{5}{c}{\textbf{Generators}} \\
        \textbf{Method} & 10 & 20 & 30 & 40 & 50 \\
        \midrule
        Deterministic & 0 & 0 & 1 & 5 & 13  \\
        Stochastic & 0 & 1 & 10 & 12 & 18 \\
        \bottomrule
    \end{tabular}
    \label{tab:dnf counts}
\end{table}

\begin{figure}[t]
    \centering
    \includegraphics[width=0.97\linewidth]{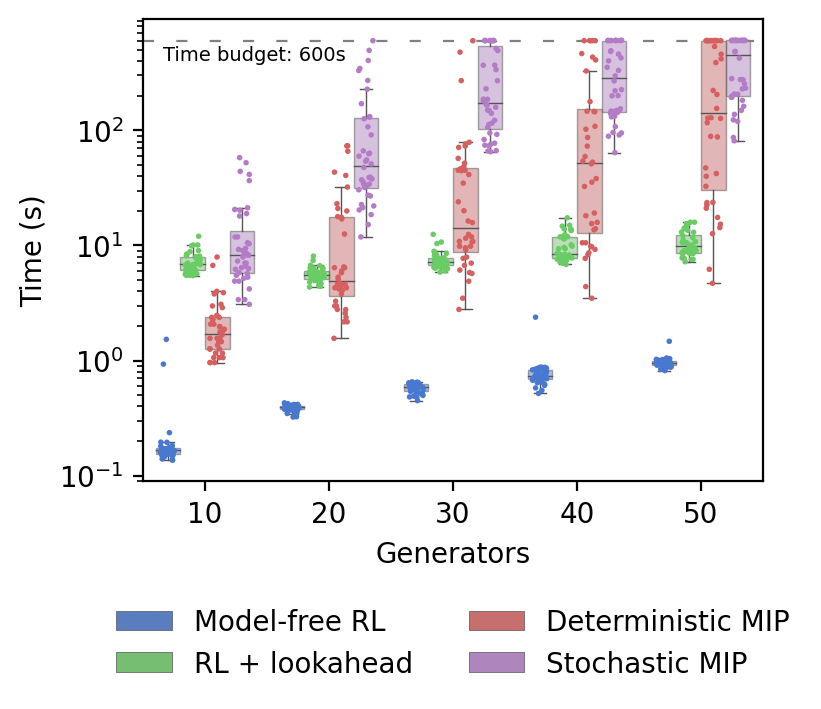}
    \caption{Distribution of solve times for the optimisation agents over the 40 case studies. The RL agents have relatively stable solve times with increasing problem size, whereas the solve times for MIP rise super-linearly with the number of generators. Solve times are limited to a time budget of 10 minutes (600 seconds).}
    \label{fig: runtime}
\end{figure}

With increasing numbers of generators, the solve times of the MIP methods increase significantly, as shown in Figure \ref{fig: runtime}. MIP solves which time out after 10 minutes do not reach the 0\% MIP gap threshold, and hence are further away from optimality than those which complete. By contrast, the run time of RL-MF exhibits linear complexity and RL-LA remains roughly constant for all problem sizes. Run times of both RL agents remain low for all problem sizes, with maximum solve times of 2s and 17s for RL-MF and RL-LA, respectively.

The number of solves (maximum 40) which did not complete within 10 minutes are shown in Table \ref{tab:dnf counts} for both MIP agents. When the MIP gap is not reached the solver outputs the current feasible solution with the lowest objective value (the current UB, as detailed in Algorithm~\ref{alg:B and B1}). For the 10 generator case, both D-MIP and S-MIP succeed in solving all 40 case studies with 0\% MIP gap. The number of unfinished solves increases with problem size for both D-MIP and S-MIP. At 50 generators nearly half of S-MIP solves do not complete, with a maximum final MIP gap of 0.23\%. Since lower MIP gaps correlate with lower cost solutions, the increasing number of unfinished solves with the number of generators indicates worsening relative solution quality as the problem size increases. However, as discussed in Section \ref{sec: cost comparison}, the deterioration in solution quality with problem size for RL agents is more much severe than for MIP agents.

\subsection{Schedule Comparison}

\begin{figure*}[t]
    \centering
    \includegraphics[width=0.97\linewidth]{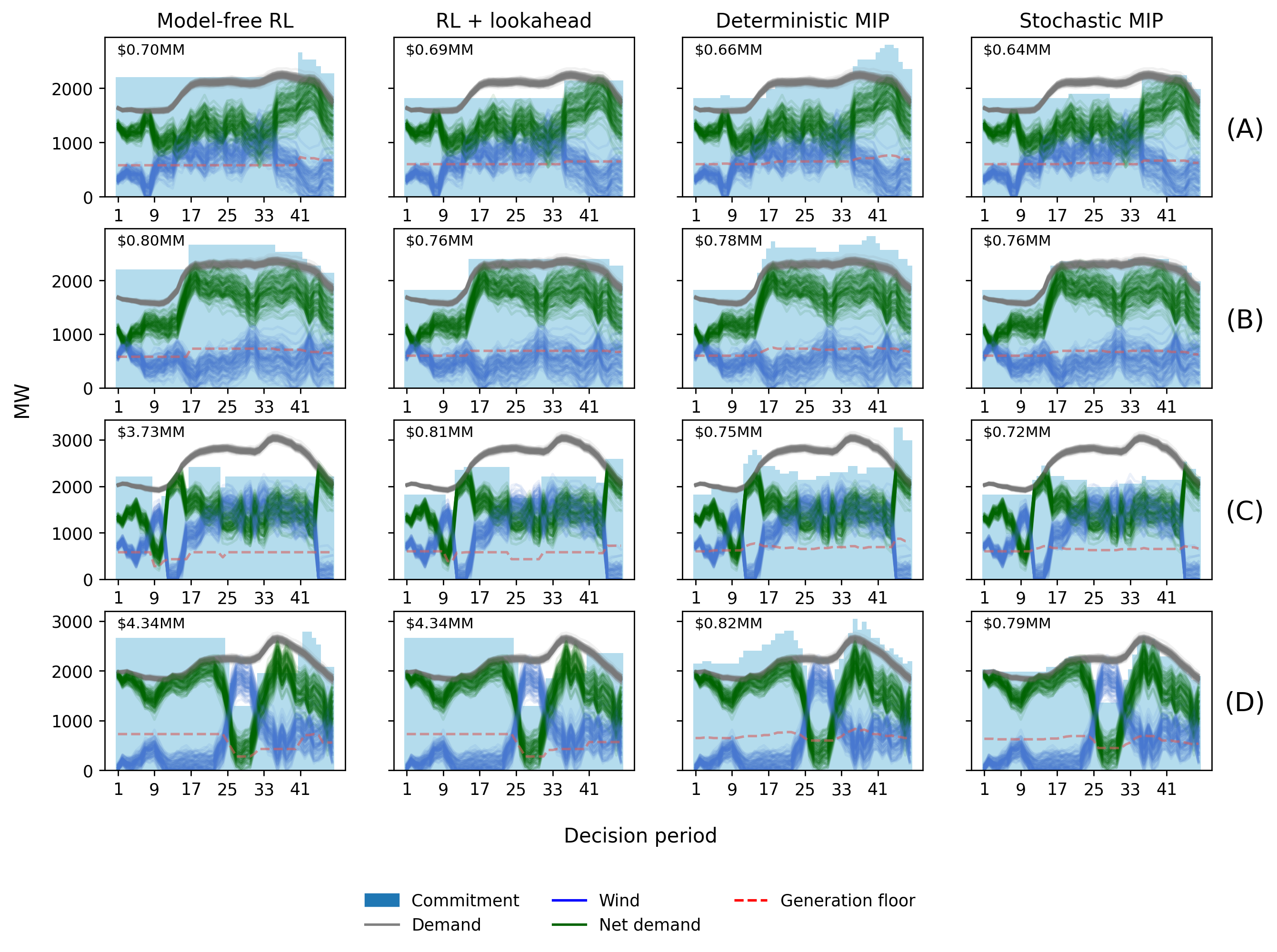}
    \caption{Selected 20 generator case studies, solved by the four optimisation agents. A subset of 100 scenarios of demand and wind are shown for each problem. In case A, both MIP methods outperform the two RL methods, with S-MIP producing the lowest operating costs overall. RL-LA outperforms RL-MF and employs small reserve margins. In case B, RL-LA outperforms D-MIP, committing less reserve. Case C is characterised by two large swings in wind generation; in this case, RL-LA is 78\% cheaper than RL. In case D, both RL-MF and RL-LA perform poorly as compared with the MIP methods, with large levels of load shedding during the evening peak.}
    \label{fig:schedule comparison}
\end{figure*}

The solutions produced by the four optimisation agents exhibit qualitative differences, as demonstrated by the 20 generator case studies in Figure \ref{fig:schedule comparison}. Case A shows a typical example of the hierarchy of operating costs, with S-MIP producing the lowest operating costs, followed by D-MIP, RL-LA and RL-MF. RL-MF is consistently the worst performing agent, exhibiting high levels of load shedding in cases C and D. In case B, RL-LA outperforms D-MIP, operating securely with less conservative reserve margins. 

The ability to consider multiple actions at each timestep gives RL-LA a significant advantage over RL-MF in case C. The two sharp decreases in wind output around decision periods 10 and 42 are managed poorly by RL-MF, resulting in high volumes of load shedding. By contrast RL-LA commits generators earlier during the morning wind drop and manages the later spike in net demand, resulting in 78\% lower operating costs as compared with RL-MF.

In case D, both RL agents decommit baseload generation during a period of high wind generation which cannot be recommitted for several hours, resulting in high levels of load shedding during the evening peak. The RL solutions are over 5 times more expensive than MIP in this case. 

In general, the RL solutions are characterised by long periods of little or no commitment changes, as shown in cases A and D. By contrast, the MIP agents follow the net demand profile more continuously. The S-MIP agent operates with the smallest reserve margins, accurately tracking the net demand profile and resulting in more efficient use of generation and the lowest operating costs.

\section{Hybridisation}
\label{sec: hybridisation}
We have shown that S-MIP achieves significantly lower operating costs than the RL agents. On the other hand, for larger system sizes S-MIP frequently utilises the full time budget of 10 minutes, whilst RL produces solutions in less than 2s. In this section, we leverage the high accuracy of S-MIP and the fast solve time of RL to produce a hybrid solution method that outperforms either individually. 

We use the trained RL agents to provide useful starting points for warm starting the MIP solver. This replaces the generic (`vanilla') initial feasible solution method used in step 1 of Algorithm~\ref{alg:B and B1} with an improved domain-specific method. This section has two parts: first, we show how more computer resource results in lower MIP gaps, which correlates with lower cost schedules; second, we show the value of concurrently producing multiple distinct solutions with sufficiently low MIP gaps, and how the RL-MF agent can assist in doing so.

To the authors' best knowledge this is the first time that RL has been used to warm-start a UC problem solve. Encouragingly we find that in doing so our method outperforms the vanilla MIP solve over the range of computer resources tested. 

\subsection{Warm Start Method Comparison}
Three different methods were used to produce initial feasible points to warm start the solver:
\begin{enumerate}
    \item `vanilla': The generic Gurobi method.
    \item RL: The pre-trained agents.
    \item `rand': Method that generates random feasible points by randomly constraining generators to be on and off throughout the day whilst respecting their constraints.
\end{enumerate}
The RL and rand methods were each used to produce 8 different initial feasible solutions for each of the 40 test days considered. This is possible since each RL policy effectively defines a probability distribution over solutions for each case study. Thus sampling produces a diverse set of intelligently selected solutions that are guaranteed to be feasible. 

\begin{figure}[t]
    \centering
    \includegraphics[width=0.97\linewidth]{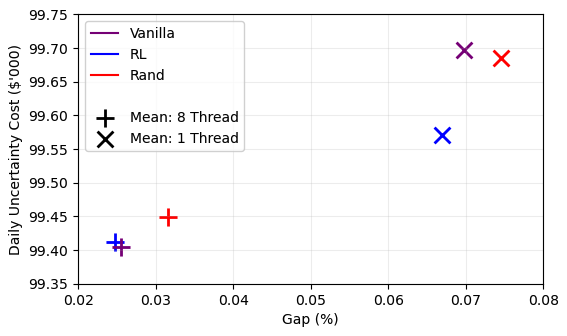}
    \caption{Average schedule cost and MIP gap over the 40 test days for the 3 warm starting methods with either 1-thread or 8-threads per solve. For the RL and rand methods, each of the 40 days was solved 8 times using 8 distinct initial solutions. More computer resources per solve result in lower average MIP gaps, which correlates with lower cost schedules when assessed via (\ref{eq: Schedule test cost}). However, this does not account for concurrency.}
    \label{fig:mean threads}
\end{figure}

Figure~\ref{fig:mean threads} plots the average MIP gap and uncertainty cost over the 40 test days when using the three warm-start methods. Uncertainty cost is the increase in a schedule's cost over that of the perfect foresight schedule's cost. For the RL and rand warm-start methods, the MIP was run 8 times for each test day, each time warm-started by a different one of the 8 distinct initial schedules. The points in Fig. \ref{fig:mean threads} are the average cost over these 8 schedules for each of the 40 test days. 

The computing resource available for each solve was either 1 or 8 threads. Comparing thread resources reveals the intuitive result that more threads available result in lower MIP gaps. This is due to the performance gains from parallelizing the solving of  concurrent leaf LPs in step 3 of Algorithm~\ref{alg:B and B1}. Lower MIP gaps are equivalent to a lower objective (\ref{eq: scenario tree cost func}), which is an approximation of the more accurate cost function (\ref{eq: Schedule test cost}) that is used to calculate the uncertainty cost on the y-axis. Hence a low MIP gap correlates with a lower uncertainty cost.

Comparison between warm-start methods utilising 1-thread reveals the first instance when hybridisation outperforms the vanilla S-MIP alone. The schedules produced from the S-MIP method with RL warm-start are, on average, \$140 or 0.14\% less expensive than the vanilla method alone. This indicates that the RL method produces better initial solutions for this problem type than the vanilla method. This improved starting point gives the RL solves an advantage, allowing them to reduce the MIP gaps faster and produce lower average cost schedules.

When 8 threads are available the vanilla and RL methods perform similarly because the increased computing power enables both methods to close the MIP gap faster. This increased speed minimises the impact of the better initial solutions offered by the RL. However, in the next section, we demonstrate how the multiplicity of solutions generated by concurrent 1-thread RL solves can be leveraged to outperform the vanilla method when all 8 threads are available.

\subsection{Concurrent Solves}
A low MIP gap indicates a lower objective value (\ref{eq: scenario tree cost func}) which is an approximation of (\ref{eq: Schedule test cost}) that is used to find the more accurate uncertainty cost of the schedules. This approximation is intrinsic to all stochastic optimisation problems because of the necessary discretisation of the continuous probability distributions. Well-constructed scenario trees will minimise but never completely eliminate this misalignment of objective functions. 

\begin{figure}[t]
    \centering
    \includegraphics[width=0.97\linewidth]{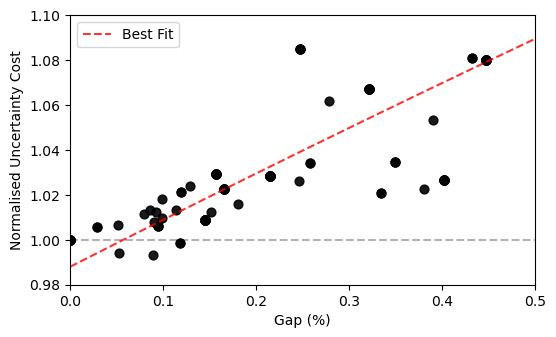}
    \caption{Cost vs MIP gap for days where the original solutions had 0\% MIP gap with 1-thread. The optimisation was re-run with varied time budgets to produce a range of final MIP gaps. This demonstrates how the misalignment of the objective being optimised (\ref{eq: scenario tree cost func}) with the true objective (\ref{eq: Schedule test cost}) results in a noisy correlation between MIP gap and a schedule's true cost, i.e. some solutions with higher gaps produce lower cost schedules.}
    \label{fig:gap vs cost}
\end{figure}

The fallout of the misalignment of (\ref{eq: scenario tree cost func}) with (\ref{eq: Schedule test cost}) is a noisy correlation between the MIP gap and a schedule's true cost. To demonstrate this, solutions with a range of MIP gap values were generated. To do this, for the 11 days where a 0\% MIP gap was achieved with 1-thread and 50 generators, the solves were re-run multiple times with time budgets ranging between 10s - 600s. The costs of these solutions were found using (\ref{eq: Schedule test cost}), normalised by the 0\% MIP gap value, and then plotted in Fig.~\ref{fig:gap vs cost}. The noisy correlation between the MIP gap and cost can be clearly observed, where although the general trend is a lower MIP gap implying lower cost, there are many instances when a higher MIP gap solution produces a lower cost schedule. Indeed, there are three instances with MIP gaps higher than 0\% where the uncertainty cost is below that of the 0\% MIP-gap schedule, demonstrating that (\ref{eq: scenario tree cost func}) is an approximation of the true objective function.

\begin{figure}[t]
    \centering
    \includegraphics[width=0.97\linewidth]{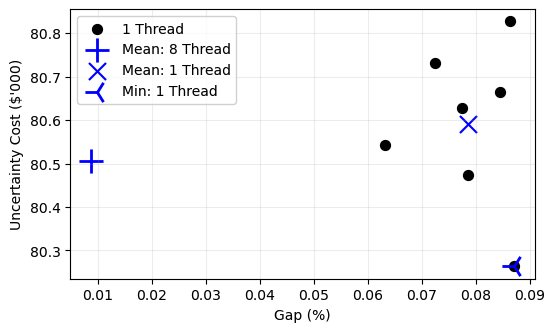}
    \caption{Concurrently warm starting multiple MIP solves with various RL solutions produces multiple low gap solutions within the time budget. Plotted here are the eight 1-thread solutions for 29/09/16. Selecting the lowest cost schedule according to (\ref{eq: Schedule test cost}) therefore exploits its slight misalignment with the MIP cost function (\ref{eq: scenario tree cost func}). This finds schedules that are lower cost than the average of 1-thread and 8-thread solves. }
    \label{fig:single day}
\end{figure}

The value of multiple concurrent solutions is illustrated in Fig. \ref{fig:single day} which plots the eight, 1-thread solutions for RL warm-start for an exemplary test day. Concurrently solving the 8, 1-thread RL and rand solutions produces 8 distinct solutions within the same low MIP gap neighbourhood. After applying (\ref{eq: Schedule test cost}) to the solutions, the cost function misalignment is exploited to give a range of costs, where the minimum (best schedule) is usually less than that of the equivalent 8-thread cost. The `vanilla' method only produces 1 initial feasible point, so cannot exploit the gains of solution multiplicity.

In Fig.~\ref{fig:single day} the solutions all have similar MIP gaps of 0.06--0.09\%, but their schedule costs range between \$80,200 to \$81,600. If concurrency was not considered, as in Fig. \ref{fig:mean threads}, then the mean 1-thread cost of \$80,720 would be expected, which is worse than the expected 8-thread cost of \$80,520. However, because all these solutions are produced within the time and computing budget, an operator can use (\ref{eq: Schedule test cost}) to choose the best solution using a more accurate metric than the optimisation's cost function or MIP gap. Thus, in this case, the operator would choose the lowest cost schedule of \$80,200, which is an improvement over the expected 8-thread solve value despite having a 0.08\% larger MIP gap.

\begin{figure}[t]
    \centering
    \includegraphics[width=0.97\linewidth]{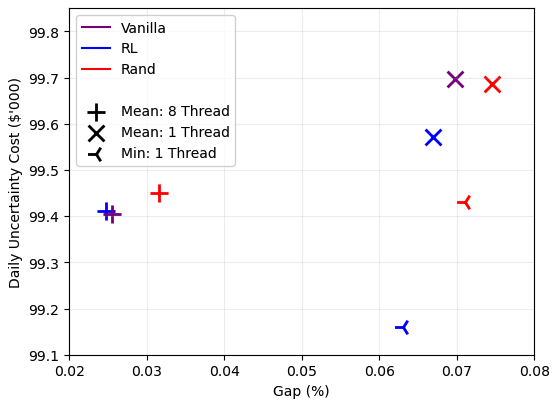}
    \caption{Taking the lowest cost, RL warm started 1-thread solve for each test day results in schedules that are on average 0.3\% lower cost than only solving a single 8-thread MIP using vanilla or RL warm start methods.}
    \label{fig:Minoverthreads}
\end{figure}

Exploiting this multiplicity of solutions for each of the 40 test-days results in an average uncertainty cost over test days of \$99,140, a reduction of 0.3\% over the previous best solutions of an individual 8-thread solve using the vanilla or RL warm-start method. These results are plotted in Fig.~\ref{fig:Minoverthreads}.

In summary, we propose a hybrid scheduling method where the fast solve times of RL are leveraged to produce a variety of distinct schedules that are used to warm-start the highly accurate S-MIP. This method produces lower cost schedules than the incumbent `vanilla' S-MIP over the range of computer resources considered. At 1-thread it produces lower MIP gap solutions, which on average perform better. At 8-thread available resources, the RL warm-start method creates value by producing a variety of distinct initial solutions, which are used to concurrently produce a variety of distinct low MIP gap solutions by the solver. So long as the final MIP gaps are sufficiently low, the randomness introduced by the misalignment of the MIP cost function (\ref{eq: scenario tree cost func}) and the more accurate cost function (\ref{eq: Schedule test cost}) can be exploited by choosing the lowest cost schedule out of the 8. This reduces average operating costs when compared to the single 8-thread solve. 

This shows that RL has good potential as a principled way of producing feasible initial solutions for the S-MIP solver, offering a significant improvement over random warm-start. To the best of our knowledge, this is the first time that RL solutions have been used to warm-start, and improve the performance of a MIP solver for the UC problem. The dependence of the hybrid method's performance on the quality and variety in the initial RL solutions is a research area that should be further investigated, given the potential savings for computationally-expensive optimisation problems such as the stochastic UC.

\section{Conclusion}
\label{sec: Conclusion}
This paper has provided a comprehensive comparison of two predominant optimisation methods for a problem of critical importance in power systems, unit commitment. RL can in principle offer rigorous handling of uncertainty and fast decision-making and has conquered several long-standing challenges of artificial intelligence \cite{mnih2015human, silver2017mastering, fawzi2022discovering}. However, our results show that currently RL is outperformed decisively by MIP methods and that significant improvements are still required to achieve competitive performance. A model-based lookahead strategy outperformed the model-free approach, albeit at greater computational cost, and was competitive with deterministic MIP for small problem sizes, but remained far below the quality of stochastic MIP. 

We characterised the issues with RL in terms of 1) scalability and 2) reliability. Regarding the former, the exponential growth of the action space with the number of generators is not handled well by existing RL methods \cite{dulac2015deep}, limiting its application to large problem sizes. Whereas median operating costs of model-free RL were 4\% higher than stochastic MIP for a 10-generator problem, this increased to 44\% for 50 generators. 

Regarding reliability, we found that the performance of the RL agents varied significantly across different problem instances. In the worst case, the operating costs of model-free RL and lookahead RL were 21 times that of stochastic MIP. Such large variability and extreme worst-case costs are unacceptable for practical applications. The MIP techniques consistently produced good quality solutions, and the presence of a measurable optimality gap is a significant advantage, providing certainty to system operators. Significant advances in core RL methods capable of handling high-dimensional discrete action spaces in security-critical contexts are required for this to be a practical alternative to MIP. 


While the RL agents were not competitive with MIP, their strong performance on some problem instances indicated that they had learned insightful properties of the UC problem. Furthermore, they are able to produce a range of feasible solutions almost instantly regardless of problem size, whereas MIP solutions routinely utilised the full time budget considered. The main methodological contribution of this paper was to leverage these RL properties to improve the MIP solve through warm starting. This method reduced operating costs by 0.3\% as compared with vanilla stochastic MIP, whilst maintaining its high consistency and explainability properties. Interestingly, we showed that the improvement relies on exploiting the mismatch between the stochastic MIP cost function and the real objective by providing multiple low MIP gap solutions concurrently. The hybrid approach shows that RL has strong potential as a method for warm starting MIP solvers for UC. 

This research highlights the value of knowledge transfer between machine learning and power systems optimisation communities to further the state-of-the-art. Future collaborations are encouraged to accelerate the development of novel UC solution methods capable of handling the high levels of uncertainty and complexity that will characterise future power systems.

\appendix
\section{}
\label{appendix_UC}
This section accompanies Section \ref{sec: MIP agent} to give the quantile-based scenario probabilities and an exhaustive list of the thermal generator constraints.
\subsection{Scenario Probabilities}
Assuming that the cost function is a continuous function of the net demand forecast quantiles $q$, the quantile scenario probabilities are found via:
\begin{subequations}
\label{scenario probability}
\begin{equation}
    \phi_1 = 0.5 \bigg( \frac{q_2^2}{q_2 - q_1} \bigg)
\end{equation}
\begin{equation}
    \phi_2 = 0.5 \bigg(q_3 - q_1 - \frac{q_1^2}{q_2-q_1} \bigg)
\end{equation}
\begin{flalign}
    & \phi_n = 0.5 \left(q_{n+1} - q_{n-1} \right) && n = 3,...,\mathcal{N}-2
\end{flalign}
\begin{equation}
    \phi_{\mathcal{N}-1} = 0.5 \bigg(q_\mathcal{N} - q_{\mathcal{N}-2} - \frac{(1-q_\mathcal{N})^2}{q_\mathcal{N} - q_{\mathcal{N}-1}} \bigg)
\end{equation}
\begin{equation}
    \phi_\mathcal{N} = 0.5 \bigg(\frac{(1-q_{\mathcal{N}-1})^2}{q_\mathcal{N} - q_{\mathcal{N}-1}} \bigg)
\end{equation}
\end{subequations}
This formulation guarantees $\sum_{n \in \mathcal{N}} \phi_n = 1$. Note that the probability of a given scenario depends on its adjacent quantile values.
\subsection{Generator Constraints}
The initial generator commitment statuses are pre-specified and enforced via:
\begin{flalign}
    & \sum_{t=1}^{UT_g - UT_g^1} (u_{g,t} - 1) = 0 && \forall g \in \cG_{\textit{on}}^1 \label{eq:a1}
\end{flalign}
\begin{flalign}
    & \sum_{t=1}^{DT_g - DT_g^1} u_{g,t} = 0 && \forall g \in \cG_{\textit{off}}^1 \label{eq:a2}
\end{flalign}
\begin{flalign}
    & u_{g,1} - U_g^1 = v_{g,1} - w_{g,1} && \forall g \in \cG \label{eq:a3}
\end{flalign}
Generator minimum up and down times are enforced through:
\begin{flalign}
    & u_{g,t} - u_{g,t-1} = v_{g,t} - w_{g,t} && \forall g \in \cG, \  \forall t \in \{2,\ldots, T \}  \label{eq:a4}
\end{flalign}
\begin{flalign}
    & \sum_{i= t-UT_g + 1}^t v_{g,i} \leq u_{g,t} && \forall g \in \cG, \ \forall t \in \{UT_g, \ldots, T\} \label{eq:a5}
\end{flalign}
\begin{flalign}
    & \sum_{i= t-DT_g + 1}^t w_{g,i} \leq 1 - u_{g,t} && \forall g \in \cG, \ \forall t \in \{ DT_g, \ldots, T\} \label{eq:a6}
\end{flalign}

\section*{Acknowledgments}

This research has been supported by the UK EPSRC project ‘Integrated  Development  of  Low-Carbon  Energy  Systems’ (IDLES, Grant EP/R045518/1). The authors acknowledge the use of UCL’s Myriad High-Performance Computing cluster for this research.


\bibliography{main.bib}

\end{document}